\documentclass[12pt]{article}

\usepackage{latexsym} 
\usepackage{amssymb}  
\usepackage{epsfig}       

\newcommand{\beq}{\begin{equation}}
\newcommand{\eeq}{\end{equation}}
\newcommand{\ba}{\begin{array}}
\newcommand{\bea}{\begin{eqnarray}}
\newcommand{\ea}{\end{array}}
\newcommand{\eea}{\end{eqnarray}}

\newcommand\comment[1]{ \hbox{[{\it Comment suppressed here.}\/]} }
\newcommand\hide[1]{}

\newcommand{\Tr}{\hbox{Tr}}

\newcommand{\bx}{{\vec x}}
\newcommand{\by}{{\vec y}}

\newcommand{\bp}{{\vec p}}
\newcommand{\bq}{{\vec q}}
\newcommand{\bk}{{\vec k}}

\newcommand{\pa}{\parallel}
\newcommand{\pe}{\perp}

\newcommand{\skipover}[1]{}

\newcommand{\C}{\mathcal{C}}

\makeatletter 

\def\appendix{\par                              
    \setcounter{section}{0}                     
    \setcounter{subsection}{0}
    \renewcommand{\theequation}{\Alph{section}.\arabic{equation}}
    \renewcommand{\thesection}{Appendix \Alph{section}}
}

\def\applabel#1{\@bsphack
  \protected@write\@auxout{}%
         {\string\newlabel{#1}{{\Alph{section}}{\thepage}}}%
  \@esphack}

\def\section{
\setcounter{equation}{0}        
\@startsection {section}{1}{\z@}{-3.5ex plus -1ex minus
 -.2ex}{2.3ex plus .2ex}{\large\bf}}
\renewcommand{\theequation}{\arabic{section}.\arabic{equation}}

\def\subsection{\@startsection{subsection}{2}{\z@}{-3.25ex plus -1ex minus
 -.2ex}{1.5ex plus .2ex}{\normalsize\bf}}

\def\subsubsection{\@startsection{subsubsection}{3}{\z@}{-3.25ex plus
 -1ex minus -.2ex}{1.5ex plus .2ex}{\normalsize}}

\makeatother   

\newsavebox{\eqlabel}

\makeatletter  
\newlength{\numblen}
\newsavebox{\eqnumb}
\def\@eqnnum{\savebox{\eqnumb}{\rm (\theequation)}%
\settowidth{\numblen}{\usebox{\eqnumb}}%
\makebox[\numblen][l]{\usebox{\eqnumb}~~~\usebox{\eqlabel}}}
\makeatother   

\newenvironment{equationwithlabel}[1]{ %
  \savebox{\eqlabel}{#1}
  \begin{equation}\label{#1} }{\end{equation}} 
\newcommand{\beql}[1]{\begin{equationwithlabel}{#1}}
\newcommand{\eeql}{\end{equationwithlabel}}

\begin{document}

\title{\bf Isotropization far from equilibrium\\[1.ex]}

\author{
J\"urgen Berges\thanks{email: j.berges@thphys.uni-heidelberg.de},$\,$
Szabolcs Bors\'anyi\thanks{email: s.borsanyi@thphys.uni-heidelberg.de},$\,$
Christof Wetterich\thanks{email: c.wetterich@thphys.uni-heidelberg.de}\\[0.2cm]
{Universit\"at Heidelberg, Institut f\"ur
Theoretische Physik}\\
{Philosophenweg 16, 69120 Heidelberg, Germany}
}

\date{}

\begin{titlepage}
\maketitle
\def\thepage{}          

\begin{abstract}
\noindent
Isotropization occurs on time scales much shorter than the
thermal equilibration time. This is a crucial ingredient for the 
understanding of collision experiments of heavy nuclei or other 
nonequilibrium phenomena in complex many body systems. We discuss
in detail the limitations of estimates based on standard ``linear'' or 
relaxation-time approximations, where isotropization and thermal
equilibration rates agree. For a weak-coupling $\phi^4$-model
the relaxation-time approximation underestimates the thermal 
equilibration time by orders of magnitude, in contrast to the  
isotropization time. The characteristic nonequilibrium 
isotropization rate can be enhanced as compared to
the close-to-equilibrium value. Our results are obtained 
from the two-particle irreducible effective action, 
which includes off-shell and memory effects and does not involve
a gradient expansion. This allows us to determine the range of 
validity of a description to lowest-order in gradients, which is 
typically employed in kinetic equations.   
\end{abstract}
\end{titlepage}

\renewcommand{\thepage}{\arabic{page}}


\section{Introduction and overview}
\label{sec:intro}
Understanding the dynamics of quantum fields far away from the ground 
state or thermal equilibrium is a challenge touching many aspects of physics,
ranging from early cosmology or collision experiments with heavy nuclei
to ultracold quantum gases in the laboratory.
One of the most crucial aspects concerns the characteristic time
scales on which thermal equilibrium is approached. 
Much of the recent interest 
derives from observations in collision experiments of heavy nuclei at RHIC.  
The experiments seem to indicate the early validity of hydrodynamics, 
whereas the present theoretical understanding of 
QCD suggests a longer thermal equilibration time. 

However, different quantities effectively thermalize 
on different time scales and a complete thermalization of all 
quantities may not be necessary to explain the observations.
We have pointed out in Ref.~\cite{Berges:2004ce} that the prethermalization 
of crucial observables may occur on time scales dramatically shorter than 
the thermal equilibration time. In particular, an
approximately thermal equation of state may be reached 
after an extremely short time. From this early time on the effective 
kinetic temperature is already very close to the final equilibrium
value of the temperature and the equilibrium
relations between average pressure, energy density and temperature
hold. Beyond the (average) equation of state, a crucial ingredient
for the applicability of hydrodynamics for collision 
experiments~\cite{Heinz:2004pj} is the approximate isotropy 
of the local pressure. More precisely, the diagonal (space-like) components
of the local energy-momentum tensor have to be approximately equal.
Of particular importance is the possible isotropization far from 
equilibrium. The relevant time scale for the early validity 
of hydrodynamics could then be set by the isotropization 
time\footnote{The use of hydrodynamics is based on conservation laws,
which involve the different components of the energy-momentum tensor.
If dissipative terms can be neglected, these equations can be
closed if an isotropic pressure can be expressed as a function 
of the energy density by an equation of state. We have shown in a 
previous publication~\cite{Berges:2004ce} that the ratio between pressure and energy density settles already at a very early prethermalization time.
Therefore, isotropization sets the relevant time scale for the use of 
hydrodynamics, in the absence of large dissipative effects as suggested by current experiments at RHIC~\cite{Heinz:2004pj}.}~\cite{Berges:2004ce,Arnold:2004ti}.

We argue in this paper that isotropization generically
happens on time scales much shorter than the thermal equilibration 
time.\footnote{A possible mechanism for fast isotropization in QCD 
in terms of plasma instabilities in high-temperature gauge theories 
has been proposed in Ref.~\cite{Arnold:2004ti}. We do not consider the 
interesting topic of plasma instabilities here. For discussions of 
instabilities in a QCD plasma see 
Refs.~\cite{Mrowczynski:1993qm,Romatschke:2003ms}.
Other approaches to address fast thermalization
include Refs.~\cite{Shuryak:2003ty}.} 
For this we concentrate on a general framework for the description of 
isotropization in far-from-equilibrium quantum field 
theory~\cite{Berges:2004yj}. 
Already for a simple scalar theory with small quartic 
self-interaction we find that
there is a large separation of times between isotropization
and thermalization. We compare our results with the most common
estimates, which evaluate a relaxation time for small excitations
away from equilibrium. This so-called ``linear'' or relaxation-time
approximation~\cite{Weldon:1983jn} assumes that there is a more or
less universal characteristic rate for many out-of-equilibrium 
properties. In particular, in the relaxation-time approximation the 
characteristic rates for isotropization and thermalization agree.
In contrast, 
we find that for rather general initial conditions the relaxation-time ansatz 
fails even to give a correct 
order-of-magnitude estimate of the thermalization time $\tau_{\rm eq}$
on which the Bose-Einstein distribution is reached. 
However, it describes well the \mbox{(on-shell)} isotropization rate, 
$\tau_{\rm iso}^{-1}$, close to equilibrium
at sufficiently late times. For earlier times, we find that the characteristic 
nonequilibrium isotropization rate can exceed the close-to-equilibrium
rate for large initial anisotropies. We go beyond the
relaxation-time approximation by calculating the nonequilibrium
dynamics from the two-particle irreducible (2PI) effective action to 
three-loop order which includes scattering, off-shell and memory effects.
The nonequilibrium evolution is solved numerically without further 
assumptions. In particular, we do not apply a gradient expansion.
This allows us to determine the range of 
validity of a description to lowest-order in gradients, which is 
typically employed in kinetic equations and to motivate
the relaxation-time ansatz close to equilibrium. The earliest time for 
the applicability of the lowest-order gradient expansion is found to
be of the order of the characteristic isotropization time $\tau_{\rm iso}$.
Many of the general findings such as a large separation of time scales between 
isotropization and thermalization, as well as the shortcomings of the
relaxation-time approximation to describe the linearized dynamics
will also be present in more complex theories like QED or QCD.
However, in gauge theories particular features such as plasma instabilities
will strongly affect the relevant time scale for 
isotropization~\cite{Arnold:2004ti}. 
\begin{figure}[t]
\begin{center}
\epsfig{file=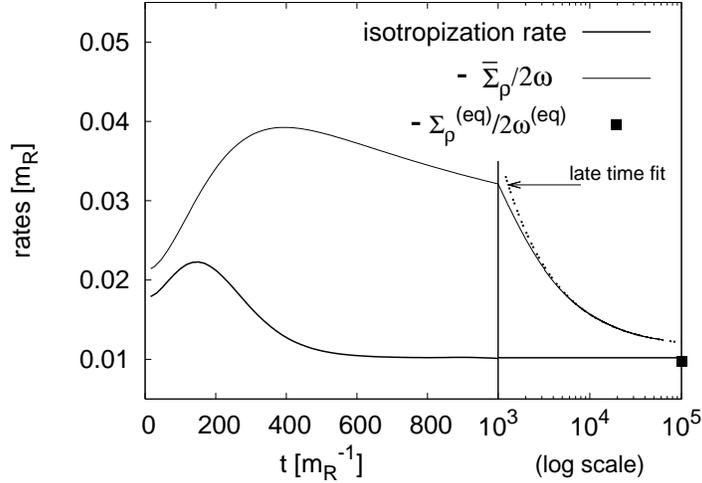,width=9.5cm}
\vspace*{-0.6cm}
\end{center}
\caption{\small Isotropization and thermalization.
The lower curve shows the characteristic nonequilibrium rate for
isotropization as a function of time $t$ in units of the renormalized
thermal mass $m_R$. At late 
times it is well approximated by the 
thermal equilibrium on-shell ratio 
$\tau_{\rm iso}^{-1} = -\Sigma^{(\rm eq)}_{\varrho}/2\omega^{(\rm eq)}$
indicated by the square symbol.
This verifies standard relaxation-time approximations for isotropization
at sufficiently late times. In contrast to isotropization, the approach to
thermal equilibrium is not well described by the
relaxation-time approximation. As an example,
the upper curve shows
$-\overline{\Sigma}_{\varrho}(t;\omega,\bp_{\rm ts})/2\omega$,
which characterizes the 
nonequilibrium evolution of the imaginary-part
of the on-shell self-energy.
Only at very late times, 
$t \gg \tau_{\rm iso} \simeq 100/m_R$,
it comes rather close to its thermal value. The dotted-line represents a
late-time fit explained in Sec.~\ref{sec:results}.
\label{fig:tworates}
}
\end{figure}

We consider a class of anisotropic initial conditions with a high 
occupation number of modes moving in a narrow momentum range around 
the ``beam direction''  $p_3  = \pm p_{\rm ts}$,  
reminiscent of some aspects of colliding wave packets in the central 
collision region (cf.\ Sec.~\ref{sec:eqs}). We compute the 
nonequilibrium
isotropization rate for a rather weak coupling of the 
$g^2 \phi^4$--interaction. 
(For our figures we use $g=1/2$ and units in terms of
the renormalized thermal mass $m_R$.) In Fig.~\ref{fig:tworates} 
we compare this rate with the relaxation-time approximation and
with the thermal equilibration time. The lower curve of the figure shows
the characteristic nonequilibrium
isotropization rate for the momentum $p_{\rm ts}$
as a function of time. For times $t \gtrsim 500/m_R$ the isotropization
rate is well described by the standard relaxation-time approximation,
i.e.\ by $\tau_{\rm iso}^{-1} =
-\Sigma^{(\rm eq)}_{\varrho}/2\omega^{(\rm eq)}$, where 
the imaginary part of the self-energy $-\Sigma^{(\rm eq)}_{\varrho}/2$
is evaluated for on-shell $\omega^{(\rm eq)}$ for momentum $p_{\rm ts}$.
The thermal equilibrium value 
$-\Sigma^{(\rm eq)}_{\varrho}/2\omega^{(\rm eq)}$ is indicated by
the full square in Fig.~\ref{fig:tworates}. At early times, one observes
that the relaxation-time rate can underestimate the characteristic 
nonequilibrium isotropization rate.

For comparison,
the upper curve of Fig.~\ref{fig:tworates} shows the 
nonequilibrium evolution of the imaginary-part
of the self-energy or, more precisely, the on-shell ratio  
$-\overline{\Sigma}_{\varrho}(t;\omega,p_{\rm ts})/2\omega$
(cf.~Sec.~\ref{sec:eqs}). Only at very late times this
rate comes rather close to its thermal equilibrium value 
$-\Sigma^{(\rm eq)}_{\varrho}/2\omega^{(\rm eq)}$. This 
reflects the slow approach of the momentum distribution
to the Bose-Einstein distribution, which will be discussed
in Sec.~\ref{sec:results}.
One observes that thermalization is an extremely slow process,
and to verify a closer approach in an actual numerical calculation 
would require to go far beyond the times shown in the figure.
Our results clearly demonstrate that the relaxation-time rate 
$-\Sigma^{(\rm eq)}_{\varrho}/2\omega^{(\rm eq)}$ cannot
be used at all to estimate the correct thermalization time. 
Most importantly, they show that isotropization
happens on time scales much shorter than the thermalization time.
This is analysed in detail in this paper.

The paper is organized a follows. In Sec.~\ref{sec:eqs} we derive the
nonequilibrium evolution equations from the 2PI effective action
and discuss the class of anisotropic initial conditions used in the
present work.
We solve the approximation of a three-loop 2PI effective action numerically,
without further approximations.  
For further analytical discussion, we consider in Secs.~\ref{sec:gradexp}
and \ref{sec:linearizeddynamics}
the linearized dynamics for small deviations from thermal equilibrium.
For this we perform a gradient expansion to lowest order and establish 
its range of applicability by comparing to the ``full'' result including 
all orders of derivatives in Sec.~\ref{sec:gradexp}. 
This is employed to derive the standard
relexation-time approximation in Sec.~\ref{sec:linearizeddynamics}.
The discussion of the numerical results is presented in Sec.~\ref{sec:results}
and we end with conclusions in Sec.~\ref{sec:conc}.

\section{Nonequilibrium evolution equations}
\label{sec:eqs}

Consider the single particle excitation which is obtained
by adding a particle with momentum $\bp$ 
to a system initially in equilibrium. For a weakly
interacting system, which has long-lived single-particle
excitations, this situation may be described in terms
of a Boltzmann equation for the distribution function $N_p(t)$ for
quasi-particles with momentum $\bp$ and energy 
$\omega_p(t)$.\footnote{Without 
loss of generality for the argument we employ here spatially homogeneous
ensembles.} 
For a small perturbation around thermal equilibrium $\delta N_p(t)
= N_p(t) - N^{\rm (eq)}_p$ the relaxation-time 
approximation yields~\cite{Weldon:1983jn,Boyanovsky:1996xx} 
\beq
\delta N_p(t) = \delta N_p(0) e^{- \gamma(\bp) t} \, ,
\eeq 
with the damping rate $\gamma(\bp)$ determined by the imaginary part of
the thermal equilibrium self-energy, 
$-\Sigma_{\varrho}^{\rm (eq)}/2$, and equilibrium on-shell frequency 
$\omega^{\rm (eq)}_p$ according to
\beq
\gamma(\bp) = - \frac{\Sigma_{\varrho}^{\rm (eq)}(\omega^{\rm (eq)}_p,\bp)}
{2 \omega^{\rm (eq)}_p} \,.
\label{eq:relaxgamma}
\eeq
This standard relaxation-time ansatz is widely applied to 
approximately describe the linearized dynamics around thermal 
equilibrium.

The question about the range of applicability of the
relaxation-time approximation has to be addressed in nonequilibrium 
quantum field theory. There are various assumptions leading to the
above interpretation of the thermal equilibrium quantity
(\ref{eq:relaxgamma}). In order to verify them we have to 
go beyond the standard gradient expansion employed in kinetic
descriptions. In addition, no quasi-particle assumption should be
employed a priori. We will therefore have to solve the nonequilibrium
time evolution including off-shell and memory effects.  

\subsection{2PI effective action and initial conditions}
\label{sec:initialcond}

Systematic approximations to describe far-from-equilibrium dynamics as well
as thermalization~\cite{Berges:2000ur,Aarts:2001qa,Cooper:2002qd,Berges:2002cz,Berges:2002wr,Juchem:2003bi,Bedingham:2003jt,Berges:2004ce,Arrizabalaga:2005tf} 
from first principles can be efficiently based
on the two-particle irreducible (2PI) effective action~\cite{Cornwall:1974vz}:
\beq
\Gamma[\phi,G] = S_{\rm cl}[\phi] + \frac{i}{2} \Tr\ln G^{-1} 
          + \frac{i}{2} \Tr\, G_0^{-1}(\phi) G 
          + \, \Gamma_2[\phi,G] \, . 
\label{eq:2PIeffact}
\eeq
To $\Gamma_2[\phi,G]$ only 
two-particle irreducible diagrams contribute, i.e.~diagrams which do
not become disconnected by opening two lines.
To be specific, we consider a scalar quantum field 
theory with classical action
\beq
S_{\rm cl}[\phi]=\int_x
	\left(\frac{1}{2}\partial^\mu \phi(x)\partial_\mu \phi(x)  
	-\frac{m^2}{2}\phi^2(x) 
	- g^2 \phi^4(x)
 \right)  \, 
\label{eq:classical}
\eeq
in the symmetric phase with vanishing field expectation
value, i.e.~$\phi = 0$. The classical inverse propagator is $i G_{0}^{-1}(x,y)=
\delta^2 S_{\rm cl}[\phi]/\delta \phi(x) \delta \phi(y)|_{\phi=0}$. 
Summation over repeated indices is implied and
we use the shorthand notation 
$\int_x \equiv \int_{\C} {\rm d} x^0 \int {\rm d}^3 x$ 
with $x \equiv (x^0,\bx)$
and $\C$ denoting a closed time path along the real axis starting at the
initial time $t = 0$. 

The equation of motion for $G$ is given by the stationarity condition 
$\delta \Gamma/\delta G = 0$, which from 
Eq.~(\ref{eq:2PIeffact}) reads~\cite{Cornwall:1974vz} 
\beq
G^{-1}(x,y) = G_0^{-1}(x,y) - \Sigma(x,y;G) \, .
\label{eq:SD}
\eeq
The self-energy $\Sigma$ is related
to $\Gamma_2[G] \equiv \Gamma_2[\phi=0,G]$ as
\beq
\Sigma(x,y;G) =   
2 i\, \frac{\delta \Gamma_2[G]}{\delta G(x,y)} \, .
\eeq 
In order to make the real and imaginary parts explicit we 
use the decomposition identity~\cite{Aarts:2001qa,Berges:2000ur} 
\beq
G(x,y) = F(x,y) - \frac{i}{2} \rho(x,y)\, {\rm sgn}_{\C}(x^0 - y^0) \, ,
\label{eq:decompprop}
\eeq
where $F(x,y)$ denotes the statistical two-point function and
$\rho(x,y)$ the spectral function. The spectral function  
encodes the equal-time commutation relations:\footnote{It is also 
directly related to the
retarded propagator $\rho (x,y) \Theta (x^0 - y^0)$, or the
advanced one $ - \rho (x,y) \Theta (y^0 - x^0)$.}
\beq
\rho(x,y)|_{x^0=y^0} = 0 \quad, \quad 
\partial_{x^0}\rho(x,y)|_{x^0=y^0} = \delta(\bx-\by) \, .
\label{eq:bosecomrel}
\eeq
To obtain a similar decomposition for the self-energy,
we separate $\Sigma$ in a ``local'' 
and ``nonlocal'' part according to
\beq
\Sigma(x,y;G) = - i \Sigma^{(0)}(x;G) \delta(x-y)
+ \Sigma^{\rm (nl)}(x,y;G) \, . 
\label{eq:sighominh}
\eeq
Since $\Sigma^{(0)}$ just corresponds to a space-time dependent 
mass-shift it is convenient for the following to introduce the notation
\beq 
M^2(x;G) = m^2 + \Sigma^{(0)}(x;G) \, .
\label{eq:localself}
\eeq
The imaginary part of the self-energy, $-\Sigma_{\rho}/2$, is determined
by 
\beq
\Sigma^{\rm (nl)} (x,y) = \Sigma_F(x,y) - \frac{i}{2} \Sigma_{\rho}(x,y)\, 
{\rm sgn}_{\C}(x^0 - y^0) \, .
\label{eq:decompself}
\eeq

Nonequilibrium dynamics requires the specification
of an initial state. While the corresponding initial conditions 
for the spectral function are
governed by the commutation relations (\ref{eq:bosecomrel}),
the statistical function $F(x,y)$ and first derivatives  
at $x^0 = y^0 = 0$ have to be specified.
Here we consider systems described by Gaussian {\em initial} density matrices.
This represents no approximation for the (non-Gaussian) dynamics
for times $t > 0$, but just constrains the class of initial
conditions. We consider a situation with an initially high occupation number 
of modes moving in a narrow momentum range around the ``beam direction''
$p_3 \equiv p_\pa = \pm p_{\rm ts}$. The occupation
numbers for modes with momenta perpendicular to this direction, 
$p_1^2 + p_2^2 \equiv p_{\pe}^2$, are small or vanishing.
The situation is reminiscent of some aspects of the anisotropic
initial stage in the central region of collision experiments of 
heavy nuclei.\footnote{Other interesting scenarios
include ``color-glass''-type initial conditions with 
distributions $\sim \exp(-\sqrt{p_\pe^2}/Q_s)$ peaked around $p_3 =0$
with ``saturation'' momentum~$Q_s$.}   
More explicitely,
we employ a class of initial conditions parametrized as 
\beq
F(t_1,t_2;\bp)|_{t_1=t_2=0} 
= \frac{n_0(\bp)+1/2}{\omega_p} \, ,
\label{eqn:init1}
\eeq
with $\partial_{t_1}F(t_1,0;\bp)|_{t_1=0} = 0$,  
$\partial_{t_1}\partial_{t_2}F(t_1,t_2;\bp)|_{t_1=t_2=0} = [n_0(\bp)+1/2]\, 
\omega_p$
for $\omega_p \equiv \sqrt{p_{\pe}^2 + p_{\pa}^2 + M_0^2}$ and $M^2_0 
\equiv M^2(t_1=0)$. The initial distribution function $n_0(\bp)$
is peaked around the ``tsunami'' momentum $p_{ts}$ with amplitude
$A$ and width $\sigma$:
\beq
n_0(\bp) = A\, \exp\left\{-\frac{1}{2\sigma^2}
\left[p_{\pe}^2 + (|p_{\pa}| - p_{\rm ts})^2 
\right]\right\} \, .
\label{eqn:init2}
\eeq
With these initial conditions, for our scalar model the most general 
statistical two-point function can be written as 
\beq
F(t_1,t_2;\bp) = F(t_1,t_2;p_{\pe},p_{\pa}) \, ,
\eeq
and equivalently for the spectral function, as well as for the self-energies
$\Sigma_{F}(t_1,t_2;\bp) = \Sigma_{F} (t_1,t_2;p_{\pe},p_{\pa})$ and 
$\Sigma_{\rho}(t_1,t_2;\bp) = \Sigma_{\rho} (t_1,t_2;p_{\pe},p_{\pa})$.

All momentum integrals appearing in this paper will be regularized on a 
lattice, with the momentum cutoff $\Lambda$ chosen such that
$\Lambda \gg p_{\rm ts}$. The renormalization of 2PI effective actions
has been discussed in detail in Refs.~\cite{Berges:2005hc,vanHees:2001ik,Blaizot:2003br,Cooper:2004rs}. For the weak couplings employed here we follow
the lines of Ref.~\cite{Arrizabalaga:2005tf} to ensure that the relevant 
length scales are larger than the lattice spacing. We note that
quantities such as the renormalized mass or renormalized damping rates 
can be directly inferred from the oscillation frequency and damping
of $F(t_1,t_2;\bp=0)$ or $\rho(t_1,t_2;\bp=0)$ 
(see Eq.~(\ref{eq:fit}) below)~\cite{Berges:2004yj}.

\subsection{Time-evolution equations}

The exact time-evolution equations for known self-energies 
$\Sigma_F$ and $\Sigma_\rho$ are obtained from the stationarity condition
of the effective action, or Eq.~(\ref{eq:SD}) by convolution with $G$.
Using the decomposition indentities (\ref{eq:decompprop}) and 
(\ref{eq:decompself}),  
these are coupled differential equations for the spectral and statistical 
functions~\cite{Aarts:2001qa,Berges:2000ur}:
\bea
\left[ \partial_{t_1}^2 + {\bp\,}^2 + M^2(t_1) \right]
F(t_1,t_2;\bp)
&\!\!=\!\!& 
- \int_{0}^{t_1}\!\! {\rm d}t'\, \Sigma_{\rho} (t_1,t';\bp)
F(t',t_2;\bp) \nonumber\\
&& + \int_{0}^{t_2}\!\!\! {\rm d}t'\, \Sigma_{F} (t_1,t';\bp)
\rho(t',t_2;\bp)\, ,\nonumber\\
\left[ \partial_{t_1}^2 + {\bp\,}^2 + M^2(t_1) \right]
\rho(t_1,t_2;\bp)
&\!\!=\!\!& 
- \int_{t_2}^{t_1}\!\! {\rm d}t'\, \Sigma_{\rho} (t_1,t';\bp)
\rho(t',t_2;\bp) \, . \label{eq:evol}
\eea
They are causal equations with integrals over the time history,
starting from the time $t_0=0$ at which the initial conditions
of Sec.~\ref{sec:initialcond} are specified.

A characteristic anisotropy measure can be chosen as
\beq
\Delta F(t_1,t_2;\bar{q}) \equiv F(t_1,t_2;p_{\pe}=0,p_{\pa}=\bar{q})
- F(t_1,t_2;p_{\pe}=\bar{q},p_{\pa}=0) \, ,
\label{eq:anisoF}
\eeq
which vanishes for the case of an isotropic correlator.
Its exact evolution equation for known self-energies
reads according to (\ref{eq:evol}):
\bea\lefteqn{
\left[ \partial_{t_1}^2 + \bar{q}^2 + M^2(t_1) \right]
\Delta F(t_1,t_2;\bar{q}) =}\nonumber\\
&-& \int_{0}^{t_1}\!\! {\rm d}t'\, \left[ \overline{\Sigma}_{\rho} 
(t_1,t';\bar{q})
\Delta F(t',t_2;\bar{q}) + \Delta \Sigma_{\rho} (t_1,t';\bar{q})
\overline{F}(t',t_2;\bar{q}) \right] \nonumber\\
&+& \int_{0}^{t_2}\!\!\! {\rm d}t'\, 
\left[ \overline{\Sigma}_{F} (t_1,t';\bar{q})
\Delta \rho(t',t_2;\bar{q}) + \Delta \Sigma_{F} (t_1,t';\bar{q})
\overline{\rho}(t',t_2;\bar{q}) \right]\, .
\label{eq:deltaF}
\eea
Here we have defined the average
\beq
\overline{F} (t_1,t_2;\bar{q})
\equiv \frac{1}{2}\left[ 
F(t_1,t_2;p_{\pe}=0,p_{\pa}=\bar{q})
+ F(t_1,t_2;p_{\pe}=\bar{q},p_{\pa}=0)
\right] \, ,
\eeq
and equivalently for the corresponding average of the
spectral function, $\overline{\rho} (t_1,t_2;\bar{q})$, and the
corresponding average self-energies 
$\overline{\Sigma}_{\rho} (t_1,t_2;\bar{q})$ and
$\overline{\Sigma}_{F} (t_1,t_2;\bar{q})$.

\begin{figure}[t]
\begin{center}
\epsfig{file=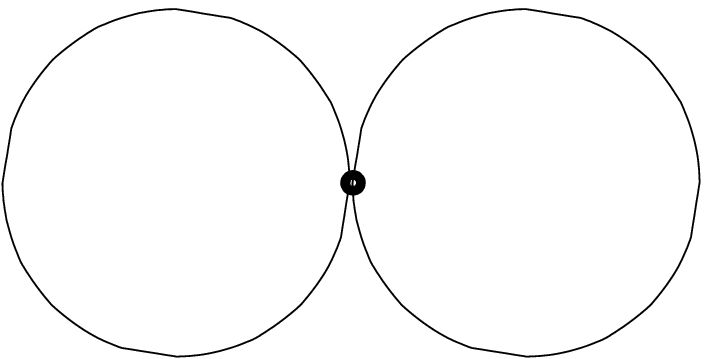,width=2.25cm} \hspace*{1cm}
\epsfig{file=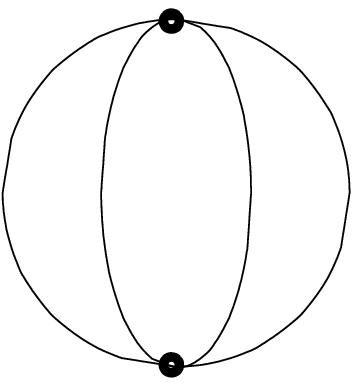,width=1.2cm}
\vspace*{-0.6cm}
\end{center}
\caption{\small
Diagrammatic representation of the contributions to $\Gamma_2[G]$ to
three-loop order.}
\label{fig:gamma}
\end{figure}

\begin{figure}[t]
\begin{center}
\epsfig{file=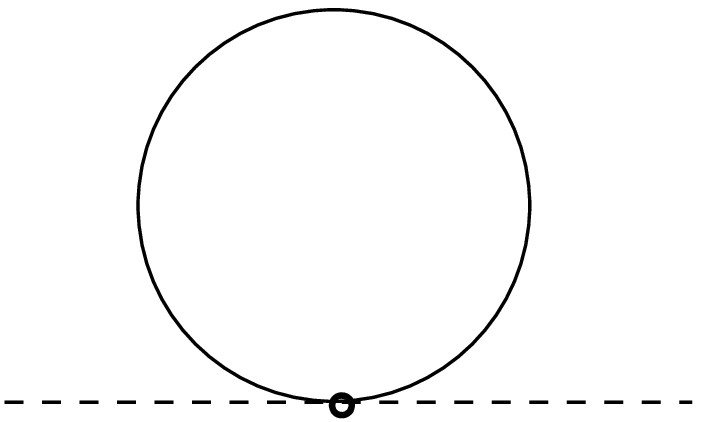,width=2.cm}\hspace*{1cm}
\epsfig{file=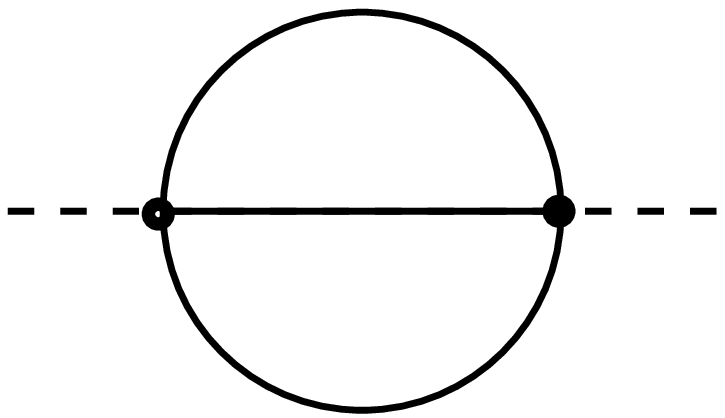,width=2.1cm}
\vspace*{-0.6cm}
\end{center}
\caption{\small
Diagrammatic representation of the corresponding contributions
to the self-energy $\Sigma$.}
\label{fig:self}
\end{figure}

We solve Eq.~(\ref{eq:evol}) numerically using the coupling- or
loop-expansion
of the 2PI effective action to order $g^4$ or three-loop order without further
approximations. To this order
the effective mass term is
\beq
M^2(t_1) = m^2 + 12 g^2 \int_{\bp} F(t_1,t_1;\bp)
\label{eq:loopM}
\eeq
and the self-energies are~\cite{Aarts:2001qa} 
\bea 
\Sigma_{F}(t_1,t_2;\bp) &\!\!=\!\!&- 96 g^4
\int_{\bq,\bk} 
F(t_1,t_2;\bp-\bq-\bk) \Big[F(t_1,t_2;\bq)F(t_1,t_2;\bk) 
\nonumber\\
&& - \frac{3}{4}\, \rho(t_1,t_2;\bq)\rho(t_1,t_2;\bk) \Big], 
\nonumber\\
\Sigma_{\rho}(t_1,t_2;\bp) &\!\!=\!\!& - 288 g^4 
\int_{\bq,\bk} \rho(t_1,t_2;\bp-\bq-\bk)
\Big[F(t_1,t_2;\bq)F(t_1,t_2;\bk) 
\nonumber\\
&& - \frac{1}{12}\, \rho(t_1,t_2;\bq)\rho(t_1,t_2;\bk) \Big] \, , 
\label{eq:2loopsigma}
\eea
using the notation $\int_\bp \equiv \int {\rm d}^3 p/(2\pi)^3$.
The diagrammatic representation of the approximation is given 
in Figs.~\ref{fig:gamma} and \ref{fig:self}. We solve
Eqs.~(\ref{eq:evol})--(\ref{eq:2loopsigma}) for the initial 
conditions numerically without further approximations
along the lines of Refs.~\cite{Berges:2000ur,Berges:2004yj}. 

In terms of these quantities the energy density $\varepsilon\equiv T^{00}$ 
and pressure components $P^i \equiv T^{ii}$ ($i=1,2,3$) 
are given by the diagonal elements of the energy-momentum 
tensor $T^{\mu\nu}$ with
\bea
\varepsilon(t_1) &=&\frac12\int_{\bq}\Bigg[\partial_{t_1}
\partial_{t_2} F(t_1,t_2;\bq)|_{t_1=t_2}
\nonumber\\
&& +\left({\bq\,}^2+m^2+ 6 g^2 \int_p F(t_1,t_1;\bp)\right) 
F(t_1,t_1;\bq)\Bigg]
\label{eq:enmom}\\
&&-\frac14\int_{\bq}\int_0^{t_1} {\rm d}t'\left[
\Sigma_{\rm F}(t_1,t';\bq)\rho(t',t_1;\bq)
-\Sigma_{\rho}(t_1,t';\bq)F(t',t_1;\bq)\right]\, ,
\nonumber
\\[0.1cm]
P^i(t_1) &=&\frac12\int_{\bq}\Bigg[\left.\partial_{t_1}\partial_{t_2}
F(t_1,t_2;\bq)\right|_{t_1=t_2}
+2 q_i^2 F(t_1,t_1;\bq)
\nonumber\\
&&-\left({\bq\,}^2+m^2+6 g^2 \int_{\bp} F(t_1,t_1;\bp)\right) 
F(t_1,t_1;\bq)\Bigg]
\label{eq:pressure}\\
&&+\frac14\int_{\bq}\int_0^{t_1} {\rm d}t' \left[
\Sigma_{\rm F}(t_1,t';\bq) \rho(t',t_1;\bq)
-\Sigma_{\rho}(t_1,t';\bq) F(t',t_1;\bq)\right] .
\nonumber
\eea
For an efficient (numerical) evaluation of these expressions it is
advantageous to replace time integrals by time derivatives of 
$F(t_1,t_2;\bp)$ using Eq.~(\ref{eq:evol}).

\subsection{Thermal equilibrium}

We finally 
note that out of equilibrium $F$ and $\rho$ are a priori {\em not} 
related by a fluctuation-dissipation relation. In contrast, 
in thermal equilibrium the periodicity (``KMS'') condition 
for the propagator, $G(x,y)|_{x^0=0}=G(x,y)|_{x^0=-i\beta}$, 
leads in Fourier space to the relation\footnote{In order to have a 
real $\varrho^{\rm (eq)}(\omega,\bp)$ we employ the Fourier-transformation
as in Eq.~(\ref{eq:wigtrans}) below, which includes an additional
factor of $-i$.}
\beq
F^{\rm (eq)}(\omega,\bp) =
\left( n_{\rm BE}(\omega) + \frac{1}{2} \right) \, 
\varrho^{\rm (eq)}(\omega,\bp) \, ,
\label{eq:fdr}
\eeq
with the Bose-Einstein distribution $n_{\rm BE}(\omega)
= 1/[\exp(\beta\omega)-1]$ for inverse temperature $\beta$.
The equivalent relation holds for the statistical ($\Sigma_F$) and spectral
($\Sigma_{\varrho}$) part of the self-energy~\cite{Berges:2000ur}. 
We emphasize that we solve
the nonequilibrium dynamics without any assumption about the
validity of a fluctuation-dissipation relation. 
In order to compare the nonequilibrium late-time results to thermal
equilibrium, we calculate in addition the thermal solution.
This can be conveniently obtained from the $\rho$-part of Eq.~(\ref{eq:evol})
along the lines of Refs.~\cite{vanHees:2001ik,Juchem:2003bi}. 
The equation for the translation invariant thermal $\rho^{\rm (eq)}(s,\bp)$ 
with $s = t-t'$ can be closed using Eq.~(\ref{eq:fdr}) to eliminate 
$F^{\rm (eq)}$ with 
\begin{equation}
\int {\rm d}s\, F^{\rm (eq)}(s;\bp)\cos(\omega s) 
=\left(n_{\rm BE}(\omega)+\frac12\right)\int {\rm d}s\,
\rho^{\rm (eq)}(s;\bp)\sin(\omega s) \, .
\end{equation}

\section{Comparison with lowest-order gradient expansion}
\label{sec:gradexp}

For further analytical discussion --- not to solve the equations ---
we consider the exact equations
(\ref{eq:evol}) in a gradient 
expansion~\cite{gradient,Danielewicz:1982kk,Chou:1984es,Calzetta:1986cq,Greiner:1998vd,Ivanov:1998nv,Wong:1996ta,Blaizot:2001nr,LMS,Prokopec:2003pj,Berges:2004yj} to
lowest order and use it to discuss the dynamics in the vicinity of thermal 
equilibrium.\footnote{The following
general analysis is not restricted to a coupling- or loop-expansion.} 
This will also allow us to recover the relaxation-time
approximation employing the corresponding additional assumptions described
below in Sec.~\ref{sec:linearizeddynamics}.
Since the range of applicability of a gradient expansion is restricted
to sufficiently homogeneous correlations at sufficiently late times, 
one cannot consider the early-time behavior in this 
case. We emphasize, however, that we are solving the nonequilibrium evolution
using Eqs.~(\ref{eq:evol}), which do not suffer from this restriction.
In particular, our results can then be used to establish whether, 
and at what time, a gradient expansion can be applied.

For sufficiently late times $t_1,t_2 \gg t_0$ 
the quantitative results turn out 
not to be notably affected by sending 
$t_0 \to -\infty$, which is required in a 
gradient expansion for practical purposes. Introducing
relative and central coordinates, $s\equiv t_1-t_2$ and 
$t \equiv (t_1+t_2)/2$, the
Wigner transformed correlators are
\bea
F(t;\omega,\bp)
&=& \int {\rm d}s\, e^{i \omega s} F(t+s/2,t-s/2;\bp)\, , 
\nonumber\\
\varrho(t;\omega;\bp)
&=& -i \int {\rm d}s\, e^{i \omega s} \rho(t+s/2,t-s/2;\bp)
\, ,
\label{eq:wigtrans}
\eea
where the factor $i$ is introduced to obtain a real Wigner-space
$\varrho(t;\omega;\bp)$. 
Note that the time integral
over $s$ is bounded by $\pm 2t$ because of the
initial-value problem with $t_1,t_2 \ge 0$.
The equivalent transformation is done for the self-energies to
obtain ${\Sigma}_{F}(t;\omega,\bp)$ and $\Sigma_{\varrho}(t;\omega,\bp)$.
In lowest-order of the expansion in derivatives of $t$ and $\omega$,
i.e.~neglecting second derivatives,
one finds from the equations (\ref{eq:evol})~\cite{Berges:2002wt,Berges:2004pu,gradient}:
\bea
2 \omega \frac{\partial}{\partial t} F(t;\omega,\bp)
&=& \Sigma_{\varrho}(t;\omega,\bp)
F(t;\omega,\bp) - {\Sigma}_{F}(t;\omega,\bp)
\varrho(t;\omega,\bp) \, ,
\nonumber\\
2 \omega \frac{\partial}{\partial t} \varrho(t;\omega,\bp)
&=& 0 \, .
\label{eq:deriv}
\eea
We note that the structure of the gradient expanded equations
is very similar to the exact equations (\ref{eq:evol}). In lowest
order of the expansion the evolution equation for the
spectral function $\varrho(t;\omega,\bp)$ becomes trivial.
A frequently employed variant of the equations (\ref{eq:deriv})
also takes into account the change of the effective mass $M^2$
of Eq.~(\ref{eq:loopM}) with respect to $t$.
This contribution 
$\sim (\partial M^2/\partial t)(\partial F/\partial \omega)$ and
the Poisson brackets entering at second order, as well as all higher orders are
neglected in the lowest-order 
expression~\mbox{(\ref{eq:deriv})}~\cite{Berges:2002wt,gradient}.
 
\begin{figure}[t]
\begin{center}
\epsfig{file=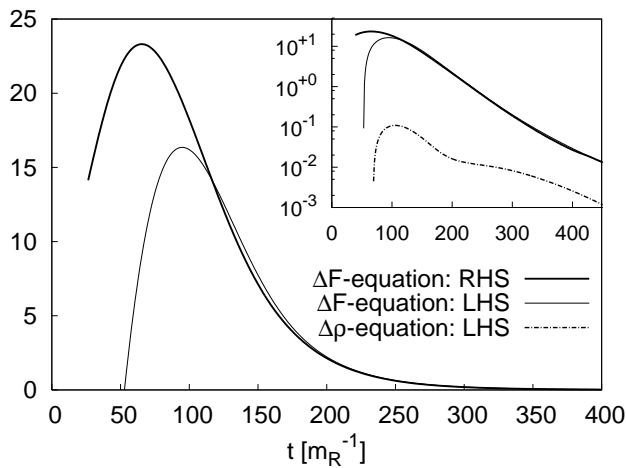,width=9.cm}
\end{center}
\caption{\small Range of validity of the gradient expansion.
The solid line shows the on-shell 
$\overline{\Sigma}_{\varrho} \Delta F + \Delta\Sigma_{\varrho} \overline{F}  
- \overline{\Sigma}_{F} \Delta \varrho - \Delta \Sigma_{F} \overline{\varrho}$
(thick line)
and $2 \omega \frac{\partial}{\partial t} \Delta F$ (thin line) in Wigner 
coordinates, as computed from 
the solution of the ``full'' equation (\ref{eq:deltaF}) without a gradient
expansion. If the gradient expansion to lowest order is correct, then
both lines have to agree according to the $\Delta F$-- and 
$\Delta \varrho$--equations~(\ref{eq:fullanisoWig}). 
The inset shows the same results on a 
logarithmic scale to make the small 
$2 \omega \frac{\partial}{\partial t} \Delta \varrho$ visible.
From these quantities we observe that the lowest-order
gradient expansion becomes valid on a time scale similar to the 
isotropization time $\tau_{\rm iso}$.}
\label{fig:wignerrel}
\end{figure}
Following the same lines, in lowest-order
of the gradient expansion one finds from Eq.~(\ref{eq:deltaF})
the anisotropy equation
\bea
2 \omega \frac{\partial}{\partial t} \Delta F(t;\omega,\bar{q})
&=& \overline\Sigma_{\varrho}(t;\omega,\bar{q})\, 
\Delta F(t;\omega,\bar{q}) 
+ \Delta \Sigma_{\varrho}(t;\omega,\bar{q})\, 
\overline{F}(t;\omega,\bar{q})
\nonumber\\ 
&-& \overline{\Sigma}_{F}(t;\omega,\bar{q})\, 
\Delta \varrho(t;\omega,\bar{q}) 
- \Delta \Sigma_{F}(t;\omega,\bar{q})\, 
\overline{\varrho}(t;\omega,\bar{q}) \, ,
\nonumber\\
2 \omega \frac{\partial}{\partial t} \Delta \varrho(t;\omega,\bar{q})&=&0
\label{eq:fullanisoWig} \, .
\eea
In order to establish the range of applicability of 
the gradient expansion, we compare with the results obtained
from Eqs.~(\ref{eq:evol})--(\ref{eq:2loopsigma}) which include {\em all}
orders of gradients: We compute the Wigner-transforms of the
results for $\Delta F(t_1,t_2;\bp)$ and 
$\Delta \rho(t_1,t_2;\bp)$ according to 
(\ref{eq:wigtrans}). From this we then evaluate
$2 \omega \frac{\partial}{\partial t} \Delta F$, and
$\overline{\Sigma}_{\varrho} \Delta F + \Delta\Sigma_{\varrho} \overline{F}  
- \overline{\Sigma}_{F} \Delta \varrho - \Delta \Sigma_{F} \overline{\varrho}$,
for on-shell $\omega$.
The results are represented as separate curves in Fig.~\ref{fig:wignerrel}.
If the gradient expansion to lowest order is correct, then
both lines have to agree according to Eqs.~(\ref{eq:fullanisoWig}), where
these contributions appear as the LHS and the RHS of the equations.
At early times the agreement
is poor and the gradient expansion to lowest order cannot be applied
as expected.  We recall that because of the initial-value problem with  
$t_1,t_2 \geq 0\,$ the time integral in the Wigner transform
(\ref{eq:wigtrans}) is finite.
However, the effects of the necessarily finite time interval for 
a nonequilibrium evolution quickly become irrelevant and do not
notably affect the curves. For times 
$t \gtrsim 100/m_R$ the two curves approach each other closely and
the full result is indeed well approximated by the lowest-order contributions
in the derivative expansion. This establishes the use of a gradient 
expansion for the anisotropy dynamics at sufficiently late times. 
We note that, in contrast to the differences 
$\Delta F$ and $\Delta \varrho$
measuring the anisotropy, the agreement was found to be much worse 
for $F$ and $\varrho$. Their description in terms of a lowest-order
gradient expansion is very poor for the times displayed here.

\section{Linear and relaxation-time approximation}
\label{sec:linearizeddynamics}

Linearized dynamics around thermal equilibrium can provide
a valid description at sufficiently late times, when the
system is approaching thermal equilibrium. 
We denote the four-momentum $p = (\omega,\bp)$ and write
\beq
F(t;p) = F^{\rm (eq)}(p) + \delta F(t;p) \, .
\eeq
Equation (\ref{eq:deriv}) can be linearized in the deviations $\delta F(t;p)$
from the \mbox{$t$-independent} thermal equilibrium value $F^{\rm (eq)}(p)$: 
\bea
\frac{\partial}{\partial t} \delta F(t;p) &=&
- \int_q  S(p,q)\, \delta F(t;q) 
\nonumber\\
& \equiv & - \int_q  S'(p,q)\, \delta F(t;q) 
+ \frac{\Sigma_{\varrho}^{\rm (eq)}(p)}{2 \omega}\,
\delta F(t;p) \, ,
\label{eq:lin}
\eea
with $\int_q \equiv \int {\rm d}^4 q/(2\pi)^4$.
Here the ``stability matrix'' $S(p,q)$ is $t$-independent and has
to be evaluated in thermal equilibrium. 
To be explicit we use the self-energies (\ref{eq:2loopsigma}) in 
thermal equilibrium 
for which the matrix $S'$ in (\ref{eq:lin}) reads 
\bea
S'(p,q) & = &   \frac{288 g^4}{\omega}  \int_k \Bigg\{
\varrho^{\rm (eq)}(p-q-k) F^{\rm (eq)}(k) F^{\rm (eq)}(p) 
\\
&-& \! \frac{1}{2} \Big[ F^{\rm (eq)}(p-q-k) F^{\rm (eq)}(k)  
+ \frac{1}{4}  \varrho^{\rm (eq)}(p-q-k) \varrho^{\rm (eq)}(k)
\Big] \varrho^{\rm (eq)}(p) \Bigg\},
\nonumber
\eea
with the equilibrium statistical and spectral function related by
Eq.~(\ref{eq:fdr}).
We have written the diagonal contribution 
$\sim \Sigma_{\varrho}^{\rm (eq)}$ explicitly in the
second line of Eq.~(\ref{eq:lin}), since the relaxation-time
approximation corresponds to assuming 
$\int S' \delta F \equiv 0$.
The omission of the ``off-diagonal'' elements in the matrix
$S'(p,q)$ is a crucial feature of the relaxation-time approximation
and responsible for its failure to describe correctly the large-time 
dynamics of thermalization as will be discussed below.\footnote{Cf.\ 
also the discussion in Ref.~\cite{Jakovac:2001kj}.} It
corresponds to the picture that the dynamics of a single excitation
$\delta F(t;p)$ can be described independently of the other modes,
assuming equilibrium values for all $F(q)$ except for the mode $q=p$.
However, the true linear description is given by the full stability matrix 
$S(p,q)$.

Using a language of discrete momenta and a 
four-volume $\Omega$, for very large time 
the evolution will be dictated by the smallest eigenvalue $\lambda_{\rm min}$
of $S/\Omega$ according to
\beq
\delta F \sim F_{\rm min} e^{-\lambda_{\rm min} t} \, ,
\eeq
with $F_{\rm min}$ the eigenvector corresponding to $\lambda_{\rm min}$.
The thermalization time can be associated with $\lambda_{\rm min}^{-1}$.
(We omit here the complications that $\lambda_{\rm min}$ may be
degenerate or vanish in the infinite volume limit.) 
The size of $\lambda_{\rm min}$ may be
much smaller than the size of typical diagonal elements of
the stability matrix or $\Sigma_{\varrho}^{\rm (eq)}/2\omega$.
The latter determines the characteristic time for a given single
mode in the relaxation-time approximation according to Eq.~(\ref{eq:isorel})
or (\ref{eq:rate}) below.
We also note that the eigenvector $F_{\rm min}$ is typically not
in the direction of the initial small deviation $\delta F$.
Deviations from thermal equilibrium will be excited for
{\em all} modes during the approach to equilibrium, explaining
the possible failure of an approximation based on a single
excitation.

In order to compute the eigenvalues of the linearization
matrix in Eq.~(\ref{eq:lin}), one may be tempted to diagonalize the matrix   
numerically from the regularized expression of the 
field-theory.\footnote{Note that the matrix is obtained from the
gradient expanded equations in the limit of an initial time in the
remote past, i.e.~$t_0 \to -\infty$. 
The latter damps 
out the unstable directions around thermal equilibrium that are 
present in the reversible dynamics for finite~$t_0$.} 
However, here we are only interested in the smallest
relevant eigenvalue which determines the late-time behavior. 
In this case, it is more efficient
to directly solve the time evolution equations (\ref{eq:evol})
along the lines of Ref.~\cite{Berges:2000ur}. The smallest eigenvalue 
(or an upper bound for it) can then
be obtained from the late-time limit of the dynamics.
In the infinite volume limit the smallest eigenvalue may approach zero.
The presence of vanishing eigenvalue(s) can invalidate the linear approach 
even for small perturbations around equilibrium. We emphasize that
the direct solution of the time evolution equations (\ref{eq:evol}) 
is not restricted to the linearized dynamics discussed in this section.

Before turning to the full numerical results in Sec.~\ref{sec:results},
we consider the dynamics in the relaxation-time approximation.
In general, the relaxation-time approximation does not describe the 
full linearized dynamics since it involves further assumptions. 
In this approximation one assumes that in the immediate vicinity of
thermal equilibrium the dynamics is well approximated by omitting
the contributions involving the matrix $S'$ in Eq.~(\ref{eq:lin}).
We first discuss isotropization. Within the relaxation-time approximation
the evolution equation for $\Delta F$ is given by 
\beq
2 \omega \frac{\partial}{\partial t} \Delta F(t;\omega,\bar{q})
\stackrel{\rm relax.\atop time}{=} \Sigma^{\rm (eq)}_{\varrho}(\omega,\bar{q})
\Delta F(t;\omega,\bar{q}) \, .
\label{eq:reltimeDF}
\eeq
The solution reads
\beq
\Delta F(t;\omega,\bar{q}) \stackrel{\rm relax.\atop time}{=}
\Delta F_0(\omega,\bar{q}) \, e^{-\gamma_{\rm iso}(\omega,\bar{q}) t} \, ,
\eeq
where the isotropization rate is determined by the 
imaginary part of the thermal self-energy, 
$-\Sigma_{\varrho}^{\rm (eq)}/2$, according 
to\footnote{Note that our results fully reproduce those obtained from 
the linearization on the level of the particle number distributions. 
The latter can simply be obtained from (\ref{eq:lin}) or (\ref{eq:reltimeDF}), 
respectively, 
by assuming a quasi-particle ansatz for the spectral function 
(i.e.\ a $\delta$-function in Wigner coordinates). Our equations are 
more general in the sense that they are not restricted to a quasi-particle 
assumption typically employed for kinetic descriptions.}
\beq
\gamma_{\rm iso}(\omega,\bar{q}) \stackrel{\rm relax.\atop time}{=}
- \frac{\Sigma_{\varrho}^{\rm (eq)}(\omega,\bar{q})}{2 \omega} \, .
\label{eq:isorel}
\eeq

We next turn to thermal equilibration.
In thermal equilibrium the ratio of the Wigner-transformed statistical
and spectral function becomes
\beq
\frac{F^{\rm (eq)}(\omega,\bp)}{\varrho^{\rm (eq)}(\omega,\bp)} = 
n_{\rm BE}(\omega) + \frac{1}{2}
\eeq
according to the fluctuation-dissipation relation (\ref{eq:fdr}),
with the \mbox{$\bp$-independent} Bose-Einstein distribution 
function $n_{\rm BE}(\omega)$.
In order to discuss the approach to thermal equilibrium, 
we consider the behavior of $F/\varrho$ at late times. In the relaxation-time
approximation one finds
\bea
\frac{F(t;\omega,\bp)}{\varrho^{\rm (eq)}(\omega,\bp)} 
&\!\!\stackrel{\rm relax.\atop time}{=}\!\!\!& 
n_{\rm BE}(\omega) + \frac{1}{2} +
\left[\frac{F_0(\omega,\bp)}{\varrho^{\rm (eq)}(\omega,\bp)} -  
n_{\rm BE}(\omega) - \frac{1}{2}
\right]\, e^{-\gamma_{\rm eq}(\omega,\bp) t} .\quad
\eea
Here we have used the fact that in thermal equilibrium
${\Sigma}_{F}^{\rm (eq)}/\Sigma_{\varrho}^{\rm (eq)} = 
n_{\rm BE} + \frac{1}{2}$.
This yields the familiar result that the thermalization rate 
in the relaxation-time approximation is determined by the 
imaginary part of the thermal self-energy according to
\bea
\gamma_{\rm eq}(\omega,\bp) \stackrel{\rm relax.\atop time}{=}
- \frac{\Sigma_{\varrho}^{\rm (eq)}(\omega,\bp)}{2 \omega} \, .
\label{eq:rate}
\eea
Of course, comparing to Eq.~(\ref{eq:isorel}) one observes the well-known 
result that the isotropization and thermalization rates agree in the 
relaxation-time approximation, i.e.\
$\gamma_{\rm iso} = \gamma_{\rm eq}$ in this 
case. We emphasize that in general isotropy is a necessary condition
for thermal equilibrium while the reverse is not. Below we will
show that beyond the relaxation-time approximation $\Delta F$ can vanish 
with a characteristic inverse rate, which is very different from the 
time for the approach of $F$ to equilibrium.

For sufficiently weak coupling the relaxation-time rate 
can be computed from perturbation 
theory, which we denote by $\gamma^{\rm (pert)}$. 
For the zero-momentum mode this is found to be~\cite{Jeon:1992kk}
\beq
\gamma^{\rm (pert)} = 
\frac{9 g^4 T^2}{2 \pi^3 m_R}\,\mbox{Li}_2(e^{-m_R/T})
\label{eq:ratepert}
\eeq
where $\mbox{Li}_2(z)$ is the second poly-logarithmic function, defined by 
\beq
\mbox{Li}_2(z)=-\int_0^{z} dw
\frac{\ln\left({1-w}\right)}{w}.
\eeq 
In the high-temperature limit this becomes
\beq
\gamma^{\rm (pert)} = \frac{3 g^4 T^2}{4 \pi m_R}.
\label{highTdr}
\eeq

Our numerical results explained in the next section show that for late 
times the isotropization rate $\gamma_{\rm iso}$ is actually well
described (on-shell) by the relaxation-time approximation. In particular,
for sufficiently 
weak coupling the perturbative expressions for the zero-momentum
mode provide a rather good estimate for this rate. On the other hand,
thermalization occurs for times that exceed by several orders of
magnitude the relaxation-time estimate (\ref{eq:rate}).    
Nevertheless, examples in the literature show that for some very specific
initial conditions the relaxation-time approximation can give a 
rather good estimate for the thermal equilibration rate as well (cf.,
in particular, Refs.~\cite{Borsanyi:2000pm,Juchem:2003bi}).

One may therefore ask what are the conditions for the
relaxation-time approximation to hold. First, one has to distinguish
between quantities whose vanishing corresponds to an enhanced
symmetry from more generic quantities. Once an ensemble is
isotropic it always remains so at late times. The vanishing of
$\Delta F$ obviously corresponds to an enhanced symmetry,
i.e.\ rotation symmetry. The isotropic states may be considered 
as a submanifold of all states: Isotropization describes then the 
time evolution for excitations orthogonal to this submanifold.
The relevant piece in the stability matrix $S$ concerns the
non-singlet representations of the rotation group. One concludes
that isotropization is determined by the smallest eigenvalue in this
submatrix rather than the smallest eigenvalue of $S$. (The stability
matrix is block diagonal in the different irreducible representations
of the rotation group.) Eigenvalues relevant for isotropization
may all be much larger than the smallest eigenvalue $\lambda_{\rm min}$, 
which typically occurs in the singlet sector, such that the
isotropization time is indeed much shorter than the thermalization
time. After the system has become isotropic there still remains a
slow flow within the isotropic subspace towards the equilibrium point. 
 
We next ask under what circumstances the relaxation-time approximation
can be used to describe thermalization. Let us start with a specific
initial condition where only one mode is excited infinitesimally.
In general, due to the off-diagonal elements in the stability matrix $S$ the
other modes will be excited as well in the course of the linear evolution. 
This defines a characteristic time scale where these
other modes reach amplitudes of the same order of magnitude as
the initially perturbed mode. The relaxation-time approximation
typically becomes invalid for larger times. On the other hand, 
for practical purposes the deviation from equilibrium may be
already substantially reduced at that time and the following
evolution of not much relevance. In this case the relaxation-time 
approximation can be used in practice. It has been actually observed
in Refs.~\cite{Borsanyi:2000pm,Juchem:2003bi} that the
relaxation-time approximation provides a rather good description
for a single initially excited mode of finite amplitude, with all 
other modes in equilibrium.
Since the off-diagonal elements of $S'$ involve a volume factor
$\Omega^{-1}$ from the \mbox{$q$-integration} in Eq.~(\ref{eq:lin}),
the relaxation-time approximation becomes valid in the infinite
volume limit if a set of modes of measure zero is initially excited.
However, realistic initial conditions involve some distribution
of initially excited modes (for example a Gaussian). Then the
sum over the initial modes in Eq.~(\ref{eq:lin}) compensates,
at least partially, the volume factor. For our initial 
conditions described in Sec.~\ref{sec:initialcond}, even though 
chosen quite narrow in momentum space, we find that the relaxation-time
approximation fails to describe the approach to thermal equilibrium.
We believe that this failure happens for a very wide class of generic
initial conditions.

\section{Numerical results}
\label{sec:results}

In the following we present
the numerical solutions of the full equations~(\ref{eq:evol}) 
with the self-energies~(\ref{eq:2loopsigma}). 
The results are then confronted
with the estimates from the relaxation-time approximation
discussed above. We employ a rather weak coupling 
$g = 0.5$ and consider a wide range of anisotropic
initial conditions (\ref{eqn:init1})--(\ref{eqn:init2}) with amplitudes
$A = 10 - 160$. The width $\sigma$ of the initial particle number
distribution $n_0(\bp)$ of Eq.~(\ref{eqn:init2}) is chosen such that
the energy expectation value is the same for the different initial
conditions. Therefore, if the system thermalizes 
then the late-time results are determined by the same temperature.
The employed energy density corresponds to that of a thermal system with
temperature \mbox{$T=1.03\, m_R$}, where we always express dimensionful 
quantities in units of the renormalized thermal mass $m_R$.
\begin{figure}[t]
\begin{center}
\centerline{
\epsfig{file=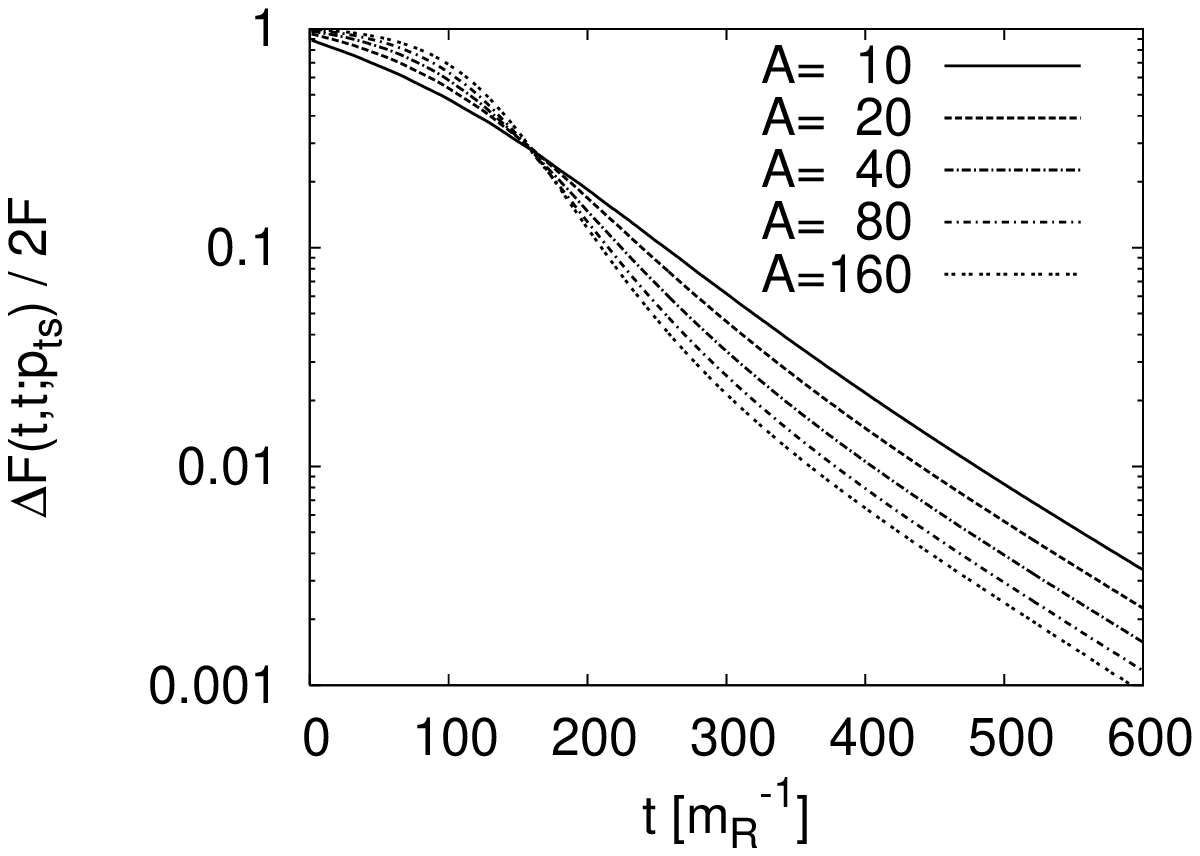,width=7.cm}
\epsfig{file=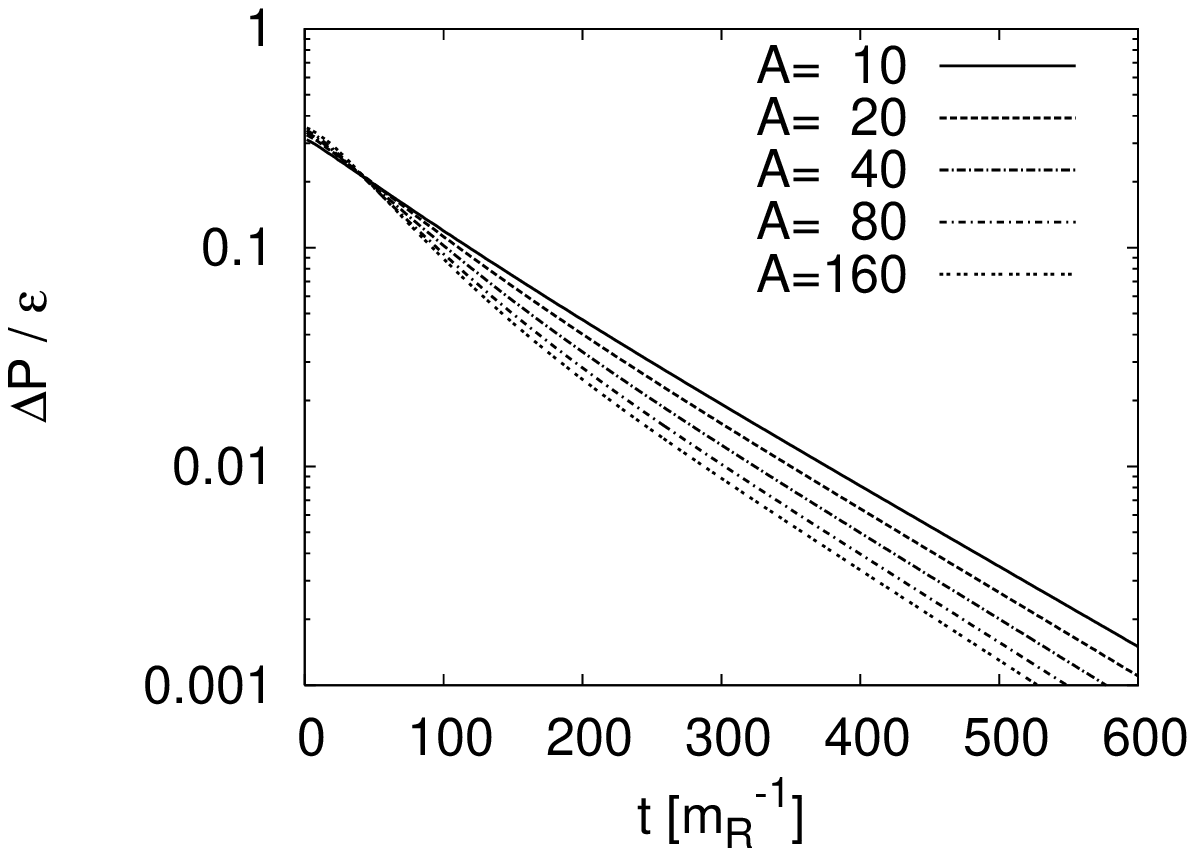,width=7.cm}
\vspace*{-0.6cm}
}
\end{center}
\caption{\small Isotropization.
On the left the normalized anisotropy correlator
$\Delta F(t,t;p_{\rm ts})$ is shown as a function of time
for different initial amplitudes $A$ with same energy density.
One observes an exponential decay of the anisotropy
with $\Delta F \to 0$ at late times.
While the behavior at early times is sensitive to the initial amplitude,
the isotropization rates become universal at late times.
On the right the corresponding pressure
differences normalized to the energy density are shown. 
\label{fig:pslopes}
\label{fig:fslopes}
}
\end{figure}

We first consider the anisotropy correlator $\Delta F(t_1,t_2;p_{\rm ts})$ 
defined in Eq.~(\ref{eq:anisoF}) with \mbox{$p_{\rm ts} = 0.93\, m_R$}. 
At equal times $t_1 = t_2$ it corresponds to the frequency integral 
of the corresponding correlator in Wigner space, i.e.\ 
$\Delta F(t,t;p_{\rm ts}) = \int [{\rm d} \omega/2\pi]\, 
\Delta F(t;\omega,p_{\rm ts})$ (cf.\ Sec.~\ref{sec:gradexp}). 
The left graph of Fig.~\ref{fig:fslopes} shows the time evolution of
$\Delta F(t,t;p_{\rm ts})$ normalized to $2 F(t,t;p_{\rm ts})$ for
different anisotropy amplitudes $A$.
One observes that at early times the behavior of $\Delta F/2 F$
depends on the initial amplitude. However, for times $t \gtrsim 300/m_R$
the isotropization rate becomes independent of the details
of the initial conditions. It approaches an exponential behavior
with rate $\gamma_{\rm iso} (p_{\rm ts}) = 0.0061(2)\, m_R$. 

A similar behavior can be seen in the right graph
of Fig.~\ref{fig:pslopes} for the pressure anisotropy 
$\Delta P (t)$
normalized to the energy density $\varepsilon$. Its late-time
isotropization rate turns out to be rather accurately described by 
the characteristic rate of the maximally populated mode, 
i.e.~$\gamma_{\rm iso} (p_{\rm ts})$. The pressure anisotropy 
and energy density have been calculated form the diagonal
components of the energy-momentum tensor (\ref{eq:enmom}) as
$\varepsilon \equiv T^{00}$ 
and $\Delta P \equiv T^{33} - T^{11} \equiv T^{33} - T^{22}$.
A value reasonable for the use of hydrodynamics, say 
$\Delta P/\varepsilon \lesssim 0.1$,
is reached particularly fast for large $A$.

In order to directly compare the
results to the relaxation-time approximation (\ref{eq:isorel}),
we consider the late-time evolution of the Wigner-space correlator 
$\Delta F(t;\omega,p_{\rm ts})$.
Following the results of Sec.~\ref{sec:gradexp}, 
$\Delta F(t;\omega,p_{\rm ts})$ 
is described by the evolution equation (\ref{eq:fullanisoWig})
for sufficiently late times.
If the relaxation-time approximation is valid then Eq.~(\ref{eq:fullanisoWig})
can be replaced by the simpler expression of Eq.~(\ref{eq:reltimeDF}).
Comparing the two equations one observes that in this case the ratio 
$-(\overline\Sigma_{\varrho} \Delta F + \Delta \Sigma_{\varrho} 
\overline{F} - \overline{\Sigma}_{F} 
\Delta \varrho - \Delta \Sigma_{F} \overline{\varrho})/2\omega \Delta F$
must approach the relaxation-time rate 
$\Sigma_{\varrho}^{\rm (eq)}/2\omega^{\rm (eq)}$.
To establish the equivalence, we first evaluate the former expression
using the full solution of the nonequilibrium evolution from
Eqs.~(\ref{eq:evol}) and~(\ref{eq:2loopsigma}). In a second calculation,
we evaluate $\Sigma_{\varrho}^{\rm (eq)}/2\omega^{\rm (eq)}$ directly
in thermal equilibrium for the same energy density. 
The results of these calculations are shown in Fig.~\ref{fig:tworates}
for on-shell $\omega$, i.e.\ evaluated from the peak of the spectral 
function~\cite{Aarts:2001qa} for momentum $p_{\rm ts}$.
The lower curve of Fig.~\ref{fig:tworates}
shows the nonequilibrium time evolution of the ratio 
$-(\overline\Sigma_{\varrho} \Delta F + \Delta \Sigma_{\varrho} 
\overline{F} - \overline{\Sigma}_{F} 
\Delta \varrho - \Delta \Sigma_{F} \overline{\varrho})/2\omega \Delta F$.
One observes that indeed for times $t \gtrsim 500/m_R$ the nonequilibrium
ratio approaches closely $-\Sigma^{(\rm eq)}_{\varrho}/2\omega^{(\rm eq)}$,
where the latter value is indicated by the square symbol on the right of the
figure. 
This verifies standard relaxation-time approximations for the description
of isotropization at sufficiently late times.
\begin{figure}[t]
\begin{center}
\epsfig{file=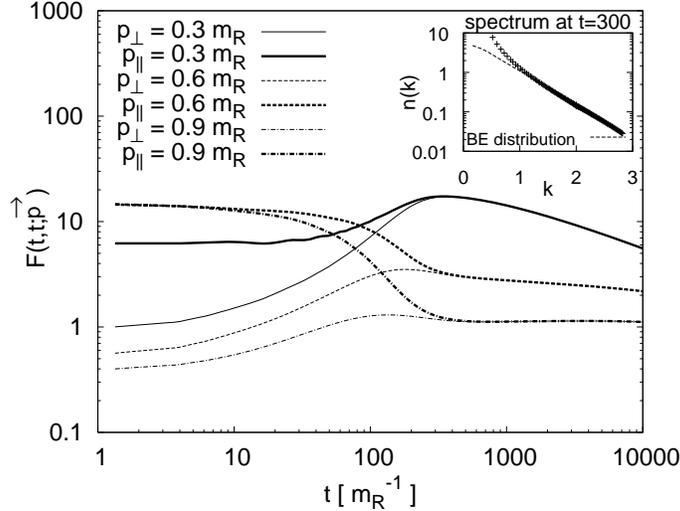,width=9.cm}
\vspace*{-0.6cm}
\end{center}
\caption{\small
Equal-time correlator $F(t,t;\bp) \equiv F(t,t;p_\pe,p_\pa)$ 
for different three-momenta
as a function of time in units of the renormalized thermal mass $m_R$.
The lines show the modes for transverse, $F(t,t;p_\pe,0)$, and
for longitudinal momenta, $F(t,t;0,p_\pa)$, for the values of
$p_\pe$ and $p_\pa$ as indicated in the figure. One observes isotropy
with $F(t,t;p_\pe,0) = F(t,t;0,p_\pa)$ for $p_\pe = p_\pa$ to
good accuracy for times $t \gtrsim 300/m_R$. In contrast,
equilibrium has not been achieved at that time and the
effectively isotropic modes continue to evolve towards a thermal 
distribution. For comparison, the dashed line of the
inset shows the Bose-Einstein (BE)
distribution along with the nonthermal time-dependent distribution 
function $n(|\bk|)$ as a function of $|\bk|/m_R$ at time $t = 300/m_R$. 
\label{fig:fiso_anizo4h}
}
\end{figure}

It should be stressed that this agreement of the nonequilibrium
isotropization rate with the normalized imaginary part of the 
{\em equilibrium self energy}, 
i.e.~$-\Sigma^{(\rm eq)}_{\varrho}/2\omega^{(\rm eq)}$, is very nontrivial
for relatively early times $t \gtrsim 500/m_R$: The imaginary
part of the {\em nonequilibrium self-energy} at that time is still far from
equilibrium! This is demonstrated in Fig.~\ref{fig:tworates},
where the upper curve shows the normalized imaginary part of the
nonequilibrium self-energy, 
or~$-\overline{\Sigma}_{\varrho}(t;\omega,\bar{q})/2\omega(t)$.
The calculation (solid line)
was performed until $t = 50000/m_R$, and the dotted line in the figure
represents a fit.\footnote{The plotted fit is obtained using a 
power-law behavior, 
however, the available late-time range of numerical data can be fitted using 
an exponential behavior with comparable accuracy.} 
Only at very late times, i.e.\ much later than the inverse
relaxation-time rate
$t \gg -2\omega^{(\rm eq)}/\Sigma^{(\rm eq)}_{\varrho}$,
its nonequilibrium evolution comes rather close to its thermal value. 
This points out that the relaxation-time
approximation fails to give an estimate for the thermalization time.
According to the relaxation-time approximation discussed in 
Sec.~\ref{sec:linearizeddynamics}, the characteristic thermalization time 
agrees with the isotropization time 
$-2\omega^{(\rm eq)}/\Sigma^{(\rm eq)}_{\varrho}$. This is sharp contrast
with the findings represented in Fig.~\ref{fig:tworates} showing
a large separation of times between isotropization and
thermalization. We emphasize that for the considered class of initial  
conditions the large separation is also found if the coupling is 
further weakened (cf.\ also the discussion in 
Sec.~\ref{sec:linearizeddynamics}).

The faster isotropization as compared to thermalization
can also be observed from the real-time correlator modes $F(t,t;\bp)$
in Fig.~\ref{fig:fiso_anizo4h}. Isotropy implies 
$F(t,t;p_\pe,0) = F(t,t;0,p_\pa)$ for $p_\pe = p_\pa$, which
is achieved to good accuracy for times $t \gtrsim 300/m_R$.
In contrast, the modes still show strong deviations from thermal equilibrium
at that time. To emphasize this fact the inset compares the time-dependent
nonthermal effective particle number 
distribution~\cite{Salle:2000hd,Aarts:2001qa,Berges:2002wr}
$n(\bk) \simeq n(|\bk|)$ as a function of $|\bk|/m_R$ at time 
$t = 300/m_R$ with the Bose-Einstein distribution. 
\begin{figure}[t]
\centerline{
\epsfig{file=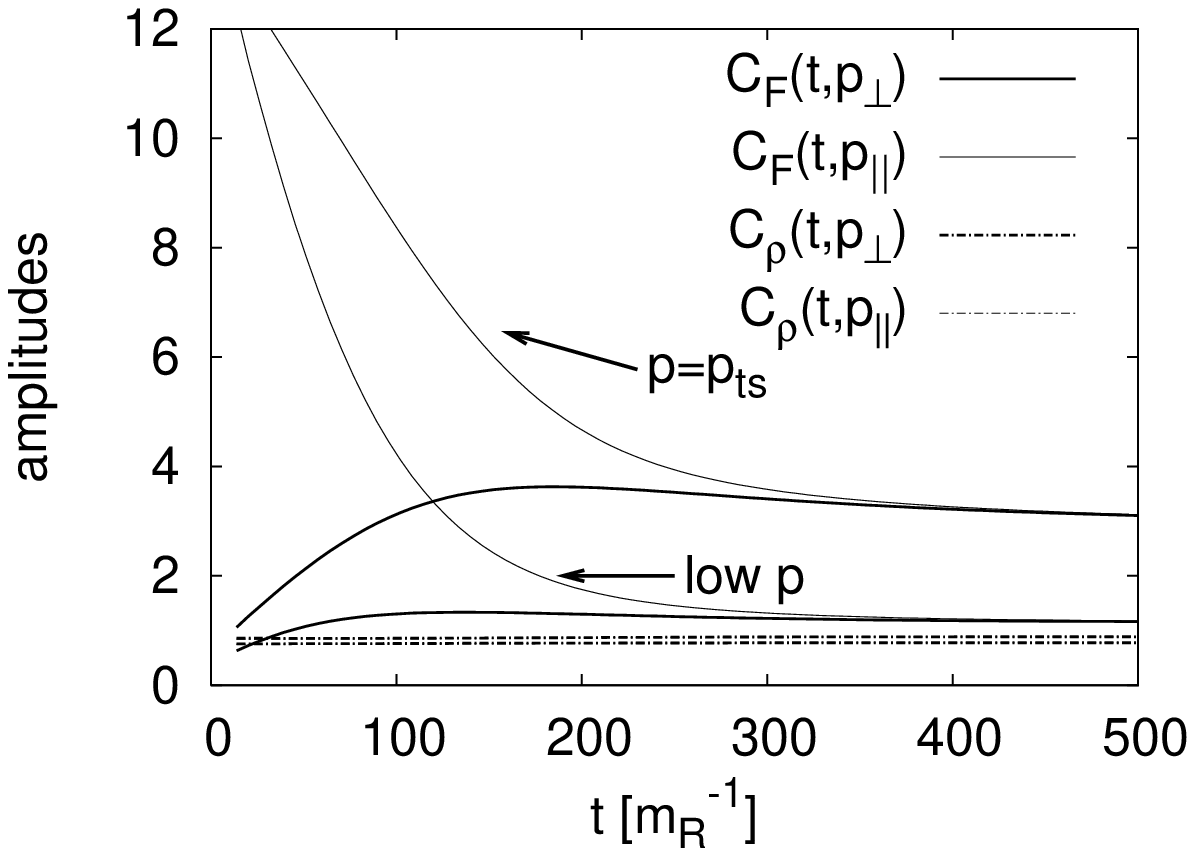,width=7.cm}
\epsfig{file=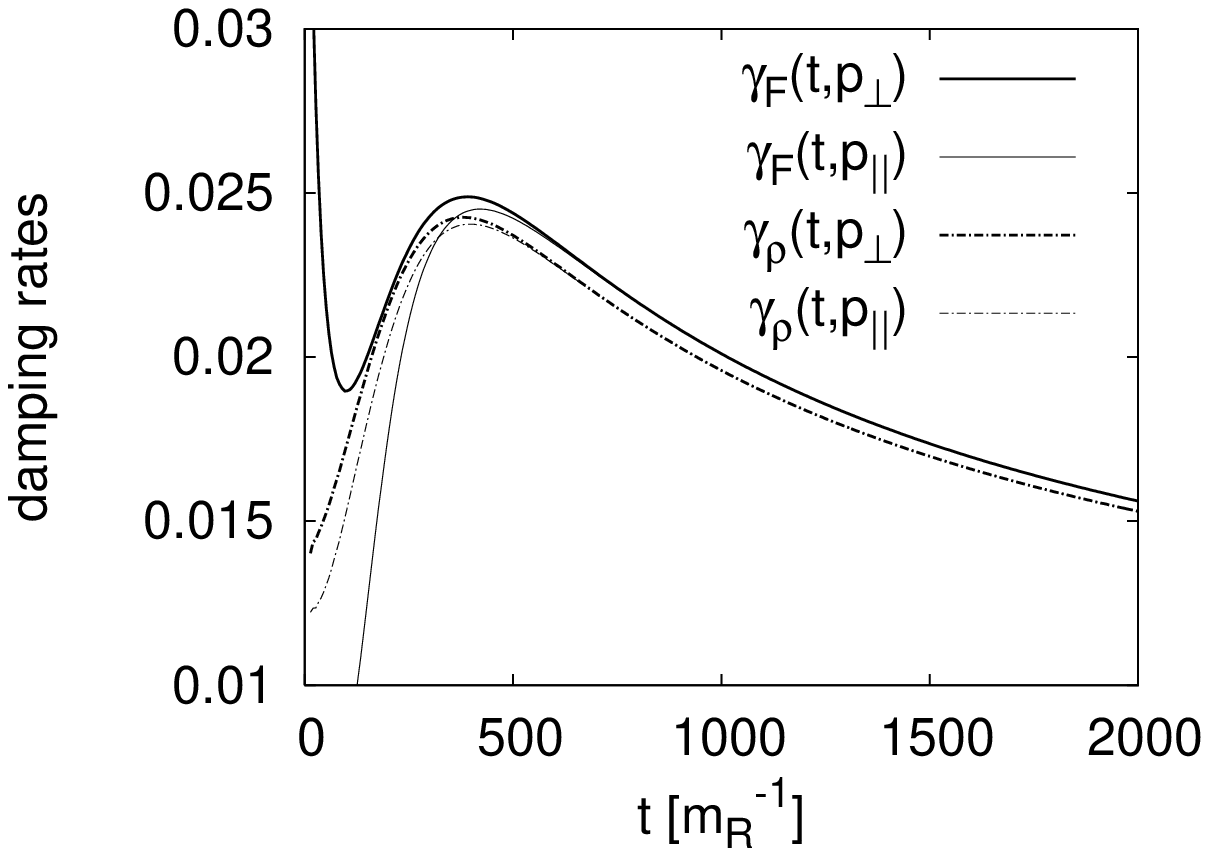,width=7.cm}
}
\caption{\small
The left graph shows the evolution of the amplitudes and the right graph shows
the damping rates extracted from the unequal-time correlators 
$F$ and $\rho$ 
as a function of $t$, as defined in Eq.~(\ref{eq:fit}) for
the maximally amplified mode with $|\bp|=p_{\rm ts}$ and for a 
low-momentum mode.
\label{fig:amplitudes}
\label{fig:dampingrates}
}
\end{figure}

The amplitudes and damping rates plotted in Fig.~\ref{fig:dampingrates} 
are the fit parameters $C_{F/\rho}$ and
$\gamma_{F/\rho}(t;\vec p)$, respectively, in
\bea
F(t_1,t_2;\bp) &=& C_{F}(t;\bp)
\cos\left[\omega_{F}(t;\bp)\,(t_1-t_2)\right]
e^{-\gamma_{F}(t;\bp)\,(t_1-t_2)}
\nonumber\\
\rho(t_1,t_2;\bp) &=& C_{\rho}(t;\bp)
\sin\left[\omega_{\rho}(t;\bp)\,(t_1-t_2)\right]
e^{-\gamma_{\rho}(t;\bp)\,(t_1-t_2)}
\label{eq:fit}
\eea
The values are extracted at each $t=(t_1+t_2)/2$ adjusting $C_{F/\rho}
(t;\bp)$, $\omega_{F/\rho}(t;\bp)$ and
$\gamma_{F/\rho}(t;\bp)$.  We
observe (not shown in the plot) an isotropic $\omega_{F/\rho}(t;\bp)$ 
parameter form the very beginning. 
In thermal equilibrium the damping parameters of Eq.~(\ref{eq:fit}) 
would equal half the
rate $-\Sigma_\varrho(t;\omega)/2\omega$ if the correlator modes
$F(t;\omega,\bp)$ and $\varrho(t;\omega,\bp)$ were of exact
Breit-Wigner form.

\section{Conclusions}
\label{sec:conc}

Isotropization happens much faster than the approach to thermal 
equilibrium. This is not a surprising fact, since only a subclass
of those processes that lead to thermalization are actually
required for isotropization: ``ordinary''
$2 \leftrightarrow 2$ scattering processes are sufficient in
order to isotropize a system, while global particle number 
changing processes are crucial to approach thermal 
equilibrium. For the massive weak-coupling scalar theory, the 
$2 \leftrightarrow 2$ processes
are of order $g^4$, while global number changing processes such
as $1 \leftrightarrow 3$ processes are of order $g^8$.  
For small couplings the $2 \leftrightarrow 2$ scatterings can, therefore, 
dominate the early-time behavior. While these processes are in principle 
sufficient to achieve complete isotropization, they are not sufficient
to provide a quantitative description of thermalization. In the absence
of global particle number changing processes they would, for instance, 
lead to a spurious chemical potential in the Bose-Einstein distribution 
at late times, which is clearly absent for the thermal equilibrium theory of 
real scalar fields.    

We demonstrate that these apparently small higher order contributions 
lead to quantitatively 
important corrections for general nonequilibrium situations. 
To establish this requires to go beyond the relaxation-time approximation,
which fails to describe the dynamics close to equilibrium in general. 
We emphasize 
in Sec.~\ref{sec:linearizeddynamics} that the relaxation-time ansatz
neglects, in particular, all off-diagonal elements in the stability 
matrix around thermal equilibrium. Whether the true small eigenvalues 
of the stability matrix 
play an important role for the late-time behavior depends on the
physical quantity of interest. We have shown that the isotropization rate is
at sufficiently late times well described by the relaxation-time 
approximation, though the  
latter fails even to give a correct order-of-magnitude 
estimate for the thermal equilibration time.  

Since our approach does not involve a gradient expansion,
we can determine the range of 
validity of a description to lowest-order in gradients, which is 
typically employed in kinetic equations and to motivate
the relaxation-time ansatz close to equilibrium. The earliest time for 
the applicability of the lowest-order gradient expansion is found to
be of the order of the characteristic isotropization time $\tau_{\rm iso}$.
Therefore, the gradient expansion and, consequently, the respective 
Boltzmann equation cannot be used to compute the isotropization time.
Similar problems with a lowest-order gradient expansion concern 
thermalization.

Far-from-equilibrium isotropization can be very important for the 
understanding of collision experiments of heavy nuclei.
An isotropic equation of state $P/\varepsilon$ is a crucial 
ingredient for the apparently successful application of hydrodynamic 
descriptions. We have shown in a previous publication
that the prethermalization~\cite{Berges:2004ce} of the equation of state 
to its equilibrium value is insensitive to the thermal equilibration 
time $\tau_{\rm eq}$.
The relevant time scale for the early validity of hydrodynamics
could then be set by the isotropization time,
which can be much shorter than $\tau_{\rm eq}$. 
In this context it is interesting to note that our simple scalar 
theory indicates that large initial anisotropies 
can lead to enhanced characteristic isotropization rates as 
compared to conventional relaxation-time rates.

\end{document}